# High spin polarization in epitaxial films of ferrimagnetic $Mn_3Ga$

H. Kurt,* K. Rode. M. Venkatesan, P. S. Stamenov and J. M. D. Coey

School of Physics and CRANN, Trinity College, Dublin 2, Ireland

* kurth@tcd.ie

## Abstract

Ferrimagnetic $Mn_3Ga$ exhibits a unique combination of low saturation magnetization ($M_s$ = 0.11 MA m$^{-1}$) and high perpendicular anisotropy with a uniaxial anisotropy constant of $K_u$ = 0.89 MJ m$^{-3}$. Epitaxial *c*-axis films exhibit spin polarization as high as 58%, measured using point contact Andreev reflection. These epitaxial films will be able to support thermally stable sub-10 nm bits for spin transfer torque memories.

There is a revival of interest in Heusler alloys, motivated by the extraordinary variety of electronic ground states that can be achieved by varying the number of valence electrons[1,2], composition[3-5] and atomic order in these materials. The Heusler family include semiconductors, metals and half metals[6], as well as ferromagnets, antiferromagnets and superconductors, and even perhaps compensated ferrimagnetic half metals[7] and topological insulators[8,9]. Thin films of half-metallic Heuslers with a high Curie temperature, such as $Co_2MnSi$, have been successfully used in magnetic tunnel junctions[10-13] and spin valves[14]. The magnetization of these films lies in-plane, but tetragonally-distorted Heuslers could offer high perpendicular anisotropy, necessary for thermally-stable sub-10 nm tunnel junctions and spin valves. Here we present epitaxially grown tetragonal $Mn_3Ga$ films, that exhibit a spin polarization of up to 58%, together with uniaxial anisotropy ($K_1 = 0.89$ MJ m$^{-3}$) and low magnetization ($M_s = 0.11$ MA m$^{-1}$), a combination of properties that may be ideal for tiny perpendicular spin-torque switchable elements that will allow for scalable magnetic memory and logic.

The general composition of the $L2_1$ cubic Heusler alloys is $X_2YZ$. In the perfectly-ordered state, illustrated in Fig. 1a, the Y and Z atoms occupy two interpenetrating face-centred cubic lattices, where each is octahedrally coordinated by the other, and the X atoms form a simple cubic lattice, where they are tetrahedrally coordinated by both Y and Z atoms, at the corners of a cube. The X–Y, X–X, Y–Y bond lengths are $\sqrt{3}a_0/4$, $a_0/2$ and $a_0/\sqrt{2}$, respectively, where $a_0$ is the cubic lattice parameter. The material we discuss here, $Mn_3Ga$, forms two stable crystal structures. The high temperature hexagonal $D0_{19}$ phase is a triangular antiferromagnet, easily obtained by arc melting[15,16]. The tetragonal $D0_{22}$ phase is a ferrimagnet, usually obtained by annealing the hexagonal material at 350-400 °C for 1-2 weeks[15, 17, 18]. The spin polarization at the Fermi level has been calculated to be 88 % for the tetragonal phase, which has been suggested as a potential material for spin torque applications[19]. To this end, thin films with appropriate magnetic properties were needed.

The $D0_{22}$ structure is a highly-distorted tetragonal variant of the $L2_1$ Heusler unit cell, which has been stretched by 27% along $c$ axis. The unit cell is outlined on Fig. 1a by the red line; lattice parameters of the unit cell (space group $I4/mmm$), which contains two $Mn_3Ga$ formula units, are $a = 394$ pm and $c = 710$ pm. As a result of the tetragonal structure, both $4d$ tetrahedral X-sites and $2b$ octahedral Y-sites are subject to strong uniaxial ligand fields, which lead to uniaxial anisotropy at both Mn positions. Imperfect atomic order results in some Mn population of the $2a$ Z sites, which are similarly distorted.

Generally, the Mn moment and Mn-Mn exchange in metallic alloys are very sensitive to the inter-atomic distances. Widely-spaced Mn atoms with a bond length $\geq$ 290 pm tend to have a large moment, of up to 4 $\mu_B$, and couple ferromagnetically[20]. Nearest-neighbour Mn atoms have much smaller moments, and couple antiferromagnetically when the bond length is the range 250 – 280 pm. As a result, the magnetic order is often complex. $\alpha$Mn, for example has a large cubic cell ($a$ = 890 pm) with four different manganese sites having moments ranging from 0.5 to 2.6 $\mu_B$, and a complex noncollinear antiferromagnetic structure with $T_N$ = 95 K. The magnetism of tetragonal $Mn_3Ga$ is simpler. The Y sublattice is ferromagnetically coupled, as expected from the long Y-Y bonds (391, 450 pm), but there is a strong antiferromagnetic X-Y intersublattice interaction (bond length 264 pm) which leads to ferrimagnetic order. This overcomes the X-X interactions, which are antiferromagnetic in-plane (bond length 278 pm) and ferromagnetic along the *c*-axis (bond length 355 pm). The antiferromagnetic X-X interactions are frustrated by the strong antiferromagnetic X-Y coupling, and the X sublattice remains ferromagnetic, with its moment aligned antiparallel to that of the Y sublattice. Different occupancies and magnetic moments in X and Y sites leads to a compensated ferrimagnetic structure with alternating spins on atomic planes along *c* axis (Fig. 1b). Imperfect atomic ordering of Mn and Ga may, however, produce local deviations from collinear ferrimagnetism. The Curie temperature of $Mn_3Ga$ is high. It would be much greater than 770 K were it not for the transition to the antiferromagnetically ordered hexagonal $D0_{19}$ structure at this temperature[15, 19].

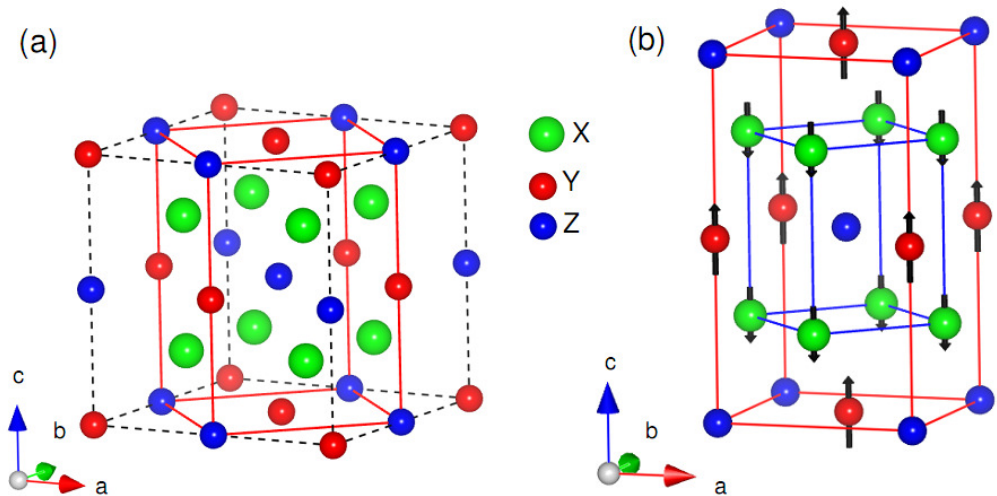

Fig. 1 (a) The cubic unit cell of the $L2_1$ Heusler alloys. The tetragonal unit cell of the $D0_{22}$ structure is indicated by the red line. (b) In $Mn_3Ga$, it is stretched along the *c*-axis. The X and Y positions are occupied by Mn, and the Z position is occupied by Ga. The magnetic couplings in X and Y sublattices are antiferromagnetic and ferromagnetic respectively, but the stronger antiferromagnetic intersublattice coupling between X and Y sites frustrates the X site moments leading to a ferrimagnetic structure.

We have grown oriented films of Mn$_3$Ga on MgO (001) substrates with various buffer layers by dc magnetron sputtering from a Mn$_3$Ga target in a chamber with $2.0 \times 10^{-8}$ Torr base pressure. The Mn$_3$Ga growth rate was approximately 1nm/min. The substrate temperature was varied from 250-375°C. Buffer layers were 10-30 nm of oriented Pt (001), Cr (001) or MgO (001). Pt seed layers were prepared by pulsed laser deposition at 500°C substrate temperature under 40 µbar oxygen partial pressure. The films grow with (001) orientation (*c*-axis normal to plane) on all three. Most of our stacks were made by dc-magnetron sputtering; Mn$_3$Ga films grown by pulsed laser deposition turned out to have the hexagonal $D0_{19}$ structure. X-ray diffraction was carried out using a Bruker D8 Discovery diffractometer with a monochromated Cu K$\alpha_1$ source. Magnetization measurements were made using a Quantum Design MPMS SQUID and PPMS vibrating sample magnetometers.

Structural, magnetic and spin polarization data for representative films are summarized in Table 1, and some X-ray diffraction data are shown in Fig. 2a-c. Sample 345 was a perpendicular spin valve structure with two Mn$_3$Ga layers, separated by a 10 nm Cr spacer. The relatively high magnetization and low spin polarization of the Mn$_3$Ga layer grown on Cr indicate that it has a high degree of disorder.

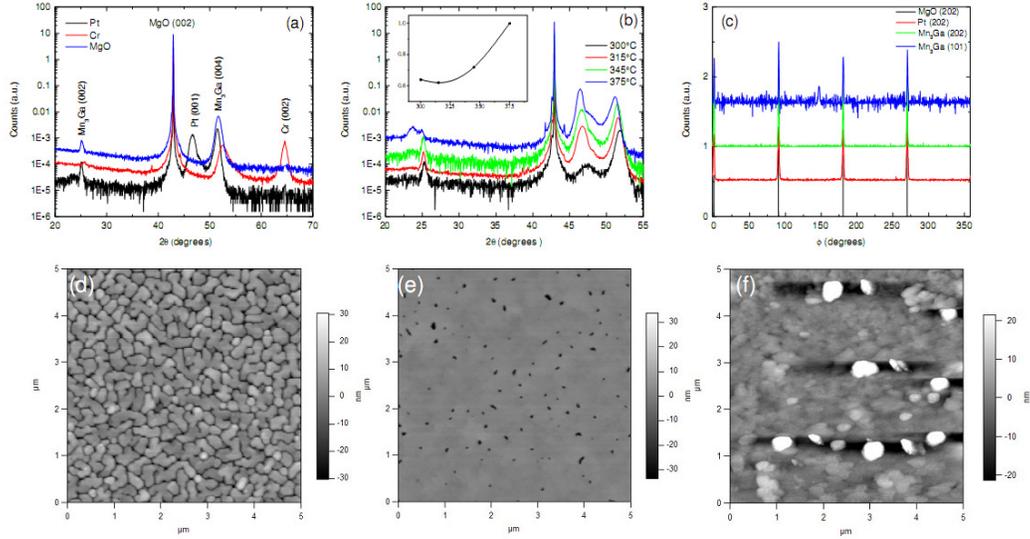

Fig. 2. a) X-ray diffraction patterns of $Mn_3Ga$ layers grown on Pt, Cr and MgO seed layers. b) Substrate temperature ($T_s$) dependence of $Mn_3Ga$ layers grown on Pt (001); the inset shows the order parameter $S$ vs. $T_s$, at $T_s = 375\,°C$ the sample is not single phase. c) $\phi$ scans showing the four-fold symmetry of the films and the MgO (001) substrate. d) Atomic force microscope images of films grown on MgO, e) Pt, f) Cr, all grown at $T_s = 350\,°C$.

Table 1. Structural, magnetic and spin polarization data on $Mn_3Ga$ films. Films grown on well oriented Pt and MgO exhibit lower magnetization accompanied with a small in-plane canted moment and higher spin polarization, whereas the films grown on Cr have higher moment and reduced polarization. Sample 345a and b represent a bilayer $Mn_3Ga$ structure; $MgO/Pt/Mn_3Ga(60)/Cr(10)/Mn_3Ga(55)$.

| Sample | Buffer | $T_s$ | $t$ | $c$ | $M\,/\!/$ | $M\perp$ | $m$ | $\mu_0 H_c$ | $P$ |
|---|---|---|---|---|---|---|---|---|---|

| # | | °C | nm | pm | (300K) MA m$^{-1}$ | (300K) MA m$^{-1}$ | (300K) $\mu_B$ fu$^{-1}$ | T | (2.2K) % |
|---|---|---|---|---|---|---|---|---|---|
| 343 | MgO/Pt | 250 | 90 | 707 | 0.022 | 0.11 | 0.65 | 1.80 | 45 |
| 346 | MgO/Pt | 250 | 90 | 712 | 0.017 | 0.10 | 0.59 | 1.90 | 55 |
| 388 | MgO | 350 | 60 | 707 | 0.015 | 0.14 | 0.81 | 1.80 | n.a. |
| 408 | MgO/Cr | 350 | 40 | 696 | 0 | 0.14 | 0.81 | 1.68 | n.a. |
| 345a | MgO/Pt | 250 | 60 | 708 | n.a. | 0.15 | 0.89 | 1.50 | 58 |
| 345b | Cr | 250 | 55 | 700 | n.a. | 0.22 | 1.30 | 1.02 | 40 |

\* 1 MA m$^{-1}$ corresponds to 5.93 $\mu_B$ fu$^{-1}$

The composition of one of the films grown on MgO was determined by ICP mass spectroscopy; the Mn:Ga atomic ratio was 3.0:1.0. The films were metallic, and the resistivity of those grown on Pt was estimated as $1.6 \times 10^{-6}$ Ω m after correcting for the resistance of the underlying Pt. Resistivity of a film grown directly on MgO could not be measured because of the discontinuous island growth mode in this case, illustrated in Fig. 2d.

An order parameter $S$ is defined by the square root of the intensity ratio of (101) to (204) reflections, divided by the theoretical ratio. The variation of $S$ with substrate temperature is shown in the inset of Fig. 2b. The highest value of 0.8 is measured for the samples grown on Pt (001) at 350 °C. Platinum has the smallest lattice mismatch with tetragonal Mn$_3$Ga (0.4%) and almost strain-free epitaxial growth takes place easily. The surfaces of Mn$_3$Ga films exhibit variations depending on the substrate type and growth temperature as shown in Fig. 2d-f. The pinhole-free areas of films grown on Pt seed layers at 350 °C have an rms surface roughness of 0.8 nm, whereas the samples grown on thin Cr seed layers were pinhole free, but much rougher (4-5 nm rms). In addition, there are large, randomly distributed islands ~ 50 nm high on the surface (Fig. 2f). Moreover, the $Mn_3Ga$ films grown on Cr have a shorter *c*-axis, which is probably caused by the large lattice mismatch with Cr (4.2 %). The $D0_{22}$ unit cell has to expand in the *ab* - plane for epitaxial growth. The films on Pt and MgO grow in cube on cube mode (Fig. 2c), whereas the films on Cr are rotated in-plane by 45 degrees to facilitate epitaxial growth. The reason for the island growth mode on MgO is thought to be the large lattice mismatch (7.7%), as islands can relax the lattice strain. The top surface of these films is smooth, but continuous film growth was not possible for thicknesses up to 60-70 nm. The pinholes seen on the surface of $Mn_3Ga$ grown on Pt simply reflect pinholes formed on the surface of the thin Pt (001) seed layer, which arise because of the lattice mismatch between Pt and MgO. Pinhole free surfaces can be obtained on thick Pt (001) seedlayers (> 200 nm) grown on MgO (001) substrates, which show featureless surfaces[21].

The relatively high surface roughness of epitaxial $Mn_3Ga$ may limit its use to giant magnetoresistance (GMR) spin valve devices. This may not be a disadvantage. As the dimensions of the memory bits decreases below 50 nm the large impedance of tunnel junctions can limit their use in high-speed circuits, whereas the impedance of GMR devices based on $Mn_3Ga$ could be engineered to 50 Ω in these dimensions, which is compatible with high-speed operation.

A typical room-temperature hysteresis loop is shown in Fig. 3a. Data are corrected for substrate diamagnetism. The substrate also exhibits weak Curie-law paramagnetism at low temperatures, which is attributed to 50 ppm of $Fe^{2+}$ impurities present in the MgO. The broad hysteresis loops show a spontaneous magnetization of 130 ± 20 kA m$^{-1}$ and coercivity of up to 1.9 T at 300 K. Epitaxial $Mn_3Ga$ films grown on MgO and Pt also exhibit a small canted magnetic moment which appears in the magnetization measurements with field parallel to the plane. There is considerable variation in coercivity, but similar values of magnetization are found in all samples, corresponding to a moment of 0.59 – 0.89 $\mu_B$ fu$^{-1}$. Values reported in bulk samples depend on the stoichiometry of the $D0_{22}$ compound, increasing from about 1.0 $\mu_B$ fu$^{-1}$ for a material with a 3:1 Mn:Ga ratio to 1.4 $\mu_B$ fu$^{-1}$ for a 2:1 ratio[16-19]. The magnetization in thin films is similar, and it has been shown to depend on the degree of atomic ordering[22, 23]. The magnetization is consistent with an essentially antiparallel arrangement of moments on the 4$d$ and 2$b$ sites, determined by neutron diffraction for a sample of composition $Mn_{2.85}Ga_{1.15}$ to be 1.6 ± 0.2 and -2.8 ± 0.3 $\mu_B$, respectively[15]. A higher moment in the 2$b$ site is consistent with the calculations of Kübler *et al.*[24]. Electronic structure calculations predict larger moments on both sites, and indicate that the compound is nearly half-metallic with a spin polarization of 88 % at the Fermi level[19].

The strength of the uniaxial anisotropy can best be determined from the perpendicular and parallel magnetization curves. The anisotropy field determined from the extrapolation in Fig. 3a is 16 T, which corresponds to a uniaxial anisotropy constant $K_1$ = 0.89 MJ m$^{-3}$. A somewhat larger value was reported by Wu *et al.* in $Mn_{2.5}Ga$, which is mainly due to the higher magnetization on account of non-stoichiometric composition[23]. The inverse correlation of coercivity and magnetization indicated in Table 1 reflects the variation of the anisotropy field $H_a = 2K_1/M_s$. The smallest volume $V$ for which the stability condition $K_u V/k_B T \geq 60$ is satisfied at room temperature is 280 nm$^3$. The corresponding nanopillar with height equal to diameter has dimensions 7 nm.

In uniformly magnetized, homogeneous uniaxial magnets, the anisotropy field sets an upper limit on the coercivity as pointed out by Stoner and Wohlfarth [25]. In reality, the coercivity is always much lower, a result known as Brown's paradox[26]. The explanation is that material is never perfectly homogeneous; reversal begins in a localized nucleation volume of $\delta_B^3$, where $\delta_B = \pi \sqrt{(A/K_u)}$ is the Bloch wall width, and $A$ is the micromagnetic exchange constant. Assuming $A \approx 10$ pJ m$^{-1}$, we find $\delta_B = 10$ nm. The expected activation volume is of order 1000 nm$^3$. We anticipate that nanostructured Mn$_3$Ga elements will exhibit stable coherent reversal at dimensions which are less than 10 nm. It therefore promises scaling to these sizes.

Multilayer spin valve stacks were realized by growing the second layer of Mn$_3$Ga on a Cr spacer. The magnetization measurements reveal a two step switching curve for the trilayer structure as shown in Fig. 3b. The magnetization reversal of the film grown on Cr is sharper with a coercivity of 1T, whereas the film grown on Pt switches at 1.5 T. The highest measured spin polarization of the Mn$_3$Ga films is 58 %, which compares with the calculated value of 88%[19]. The polarization decreases with order, strain as well as Mn deficiencies, which increases magnetization.

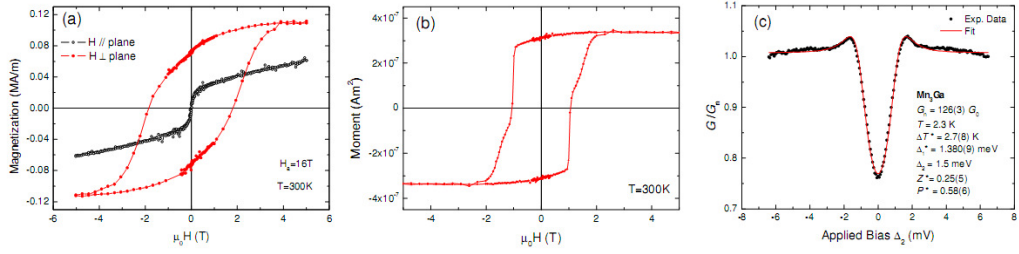

Fig. 3. (a) Magnetization curves for a 90 nm thick film of tetragonal Mn$_3$Ga measured at room temperature with the field applied perpendicular and parallel to the plane of the film. Anisotropy field at 300 K was determined by extrapolating in plane data through origin and is 16 T. (b) Magnetization of Mn$_3$Ga/Cr/Mn$_3$Ga bilayer structure. (c) Point-contact Andreev reflection spectrum of a Mn$_3$Ga film grown on Pt.

The tetragonal $D0_{22}$-Mn$_{3-x}$Ga (x = 0 - 1) offers a wide variety of magnetic properties that can be engineered to suit a specific application. In the case of stoichiometric Mn$_3$Ga, presented here, the magnetization is reduced due a higher degree of compensated spins and high anisotropy would allow thermally-stable sub-10 nm spin torque devices. For magnetic random access memories, a low magnetization has the advantage that it requires a lower current to switch by STT. A particular advantage is that the Mn$_3$Ga nanopillars would be switchable by STT but immune to magnetic field of even NdFeB permanent magnets due to the high coercivity of the films, and therefore would not require any magnetic shielding.

In conclusion, tetragonal Mn$_3$Ga thin films look promising for nanoscale spin-transfer torque memory and logic applications. There is sufficient anisotropy to ensure thermal stability to sub-10 nm dimensions, and the small magnetization is advantageous for spin torque switching. There is a degree of flexibility in the $D0_{22}$ structure, in terms of composition and degree of atomic order, which should enable the magnetic properties to be optimized for a specific magnetic application. The immediate challenge now is to observe spin torque switching a Mn$_3$Ga nanopillar.

**Acknowledgements**

This work was supported by SFI as part of the MANSE project 2005/IN/1850, and was conducted under the framework of the INSPIRE programme, funded by the Irish Government's Programme for Research in Third Level Institutions, Cycle 4, National Development Plan 2007-2013.